\documentclass[pra,floatfix,showpacs,amsmath,twocolumn]{revtex4}
\usepackage{amssymb}
\usepackage{graphicx}
\usepackage{amsfonts}
\usepackage{amssymb}

\begin{document}

\newcommand{\be}{\begin{eqnarray}}
\newcommand{\ee}{\end{eqnarray}}
\newcommand{\bea}{\begin{eqnarray}}
\newcommand{\eea}{\end{eqnarray}}
\newcommand{\bma}{\begin{subequations}}
\newcommand{\ema}{\end{subequations}}
\def\lR{l^2_{\mathbb{R}}}
\def\RR{\mathbb{R}}
\def\E{\mathbf e}
\def\D{\boldsymbol \delta}
\def\S{{\cal S}}
\def\T{{\cal T}}
\def\dd{\delta}
\def\one{{\bf 1}}
\def\kk{{\bf k}}
\def\qq{{\bf q}}
\def\rr{{\bf r}}
\def\nn{{\bf n}}
\def\LL{\textmd{L}}

\title{Quantum engineering of photon states with entangled atomic ensembles}

\author{D. Porras and J. I. Cirac}
\affiliation{Max-Planck Institut f\"ur Quantenoptik,
Hans-Kopfermann-Str. 1, Garching, D-85748, Germany}

\pacs{PACS}
\date{\today}

\begin{abstract}
We propose and analyze a new method to produce single and entangled
photons which does not require cavities. It relies on the collective
enhancement of light emission as a consequence of the presence of
entanglement in atomic ensembles. Light emission is triggered by a
laser pulse, and therefore our scheme is deterministic. Furthermore,
it allows one to produce a variety of photonic entangled states by
first preparing certain atomic states using simple sequences of
quantum gates. We analyze the feasibility of our scheme, and
particularize it to: ions in linear traps, atoms in optical lattices, 
and in cells at room temperature.
\end{abstract}

\maketitle
The deterministic generation of collimated single and entangled
photons is of crucial importance in Quantum Information, like in quantum cryptography
\cite{Gisin}, 
quantum computation \cite{KLM}, 
quantum lithography \cite{Dowling} 
or quantum interferometry \cite{Bollinger,Lloyd}.
Most of the methods tested so far require high-Q cavities, something
which is very demanding in practice \cite{Yamamoto,Rempe,Keller,Kimble.spe}. 
The engineering of quantum states in atomic systems is now
possible thanks to the experimental progress experienced by the
field of Atomic Physics during the last years. In fact, with
trapped ions it has been already possible to create so--called W
\cite{Häffner} and GHZ \cite{Leibfried} states of up to 8 ions.
At the same time, scientists have been able to produce other kinds
of entangled states \cite{Bloch} with atoms in optical lattices.
Furthermore, with the advent of Rydberg techniques
\cite{LukinRydbergatoms} it will soon be possible to create
W--like states in that system or in atomic ensembles at room
temperature. Apart from their fundamental interest, some of those
states may have applications in precision spectroscopy
\cite{Wineland,Roos}.

In this work we show that the ability of creating those atomic
states may have a strong impact in different subfields of quantum
information, as it may lead to a very efficient way of creating
certain kind of entangled photonic states which are required in
various applications. The main idea is to use
a laser and an internal level configuration such that we can map
the atomic state onto photonic states corresponding to modes
propagating in a well defined direction. Our scheme uses the well
known fact \cite{Jackson,Scully} that, under certain
circumstances, light scattering takes place predominantly in the
forward direction due to an interference effect. In fact, this
effect is the basis of one of the building blocks of the repeater
scheme proposed in \cite{DuanCiracZoller}, and has been recently
demonstrated in a series of experiments
\cite{Kimble,Lukin,Kuzmich}. There, a single excitation is created
in an atomic ensemble by detecting a photon emission in a certain
direction. Then, the excitation is released in the forward
direction by using a laser. Building on this fact, we propose to
create certain kind of excitations by using quantum gates or
atomic interactions, which give rise to the desired entangled
states when they are released using a laser, and which propagate
in the desired direction due to the mentioned interference effect.

\begin{figure}[h]
\center
\resizebox{\linewidth}{!}{\includegraphics{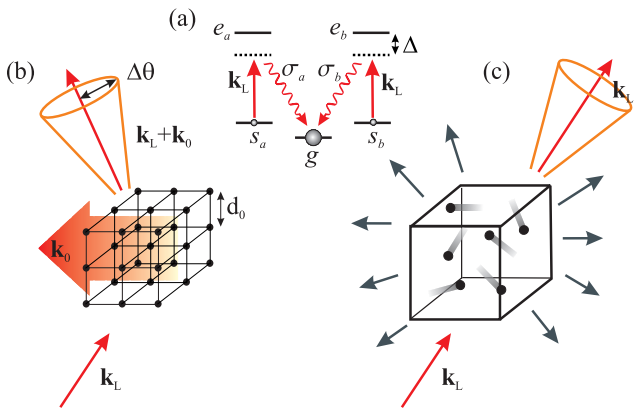}}
\caption{(a)
Level configuration for the release of atomic entangled
  states in photonic channels.
(b) Release of a collective state with linear momentum $\kk_0$, that
has been generated in a lattice of atoms.
(c) Emission of photons from an atomic ensemble, which consists of an
incoherent contribution (isotropic), and a coherent one in the
forward--scattering direction.}
\label{general}
\end{figure}

Let us consider a set of $N$ atoms with (ground) hypefine levels
$|g\rangle$ and $|s_{a,b}\rangle$ (see Fig. \ref{general} (a)). We
consider states of the form
\begin{equation}
| \kk^{(n_a)}_a , \kk^{(n_b)}_b \rangle = \frac{1}{\sqrt{n_a ! \
n_b !}} \left( \sigma^\dagger_{a,\kk_a} \right)^{n_a} \left(
\sigma^\dagger_{b,\kk_b} \right)^{n_b} | 0 \rangle ,
\label{multi.W.state}
\end{equation}
and linear combinations thereof. Here, $| 0 \rangle = | g
\rangle_1 \dots | g \rangle_N$, and
\begin{equation}
\sigma^\dagger_{x,\kk_\alpha} = \frac{1}{\sqrt{N}} \sum_{j=1}^{N}
e^{-i \kk_x \rr^0_j} \sigma_{x,j}^\dagger, \quad x=a,b,
\end{equation}
where $\sigma_{x,j}^\dagger$ excites an atom from $| g \rangle_j$
to $| s_x \rangle_j$, and $\rr_j^0$ are the equilibrium position of
the atoms.
In the limit $n_x \ll N$, Eq.
(\ref{multi.W.state}) defines a set of orthonormal collective
states with $n_x$ atoms excited in $| s_x \rangle$ and linear
momentum $\kk_x$. Those states can be indeed
readily created using trapped ions or Rydberg techniques 
(see Appendix \ref{MPS},\ref{Rydberg}).

In order to release the photons, one sends a laser pulse of
wavevector $\kk_\LL$ which couples level $|s_x\rangle$ to some
electronically excited ones $|e_x\rangle$, respectively. The large
population of level $|g\rangle$ together with the initial
entanglement (coherences) between the atoms, will now stimulate the
emission of photons from the excited states to the level $|
g\rangle $, which overall will produce the mapping between these
states and the photonic states,
\begin{equation}
| \kk^{(n_a)}_a , \kk^{(n_b)}_b \rangle \rightarrow | n_a
\rangle_{\kk_a + \kk_\LL, \sigma_a} | n_b \rangle_{\kk_b +
\kk_\LL,
  \sigma_b};
\label{mapping}
\end{equation}
that is, (\ref{multi.W.state}) is mapped to a Fock state of $n_x$
photons with momenta $\kk_x + \kk_\LL$ and polarization
$\sigma_x$, where $\sigma_x$ is the polarization of the light in
each decay channel. Moreover, due to the linearity of this
process, superpositions of states of the form
(\ref{multi.W.state}) will be mapped onto superpositions of
photonic states (\ref{mapping}). For example, the atomic state
$\left( | \kk^{(1)}, \qq^{(1)} \rangle + | \qq^{(1)}, \kk^{(1)}
\rangle \right)/\sqrt{2}$ will emit a pair of entangled photons in
different directions. 
The mapping (\ref{mapping}) is strictly valid under ideal conditions,
and in the limit $N \to \infty$, and
the directionality in the photon emission is
directly connected to the momentum conservation which, in turn, is
a consequence of the
constructive interference in the field emitted by each atom. Thus,
the crucial issue in our scheme is to determine how this mapping is
modified in finite atomic ensembles under nonideal conditions.
In the following we analyze such questions in detail,
concentrating in the simplest case in which we have a single
excitation with momentum $| \kk_0 \rangle$ in $| s_a \rangle$ 
(i.e. our initial state is a W-like state) and thus we produce a single photon. 
We determine a function $f(\Omega)$, which is proportional to
the probability density that the photon is emitted in the direction
$\Omega$. 
In general, $f = f_\textmd{coh} + f_\textmd{inc}$; that is,
it is the sum of a coherent contribution and an incoherent one.
The later appears whenever the positions of the particles
fluctuate. $f_\textmd{coh}$ contains the forward scattering
contribution, which is emitted in a cone with a width $\Delta \Omega$ that
decreases with the number of particles. $f_\textmd{inc}$, on the
contrary, describes isotropic light emission, thus, even when the light emitted in
$\Delta \Omega$ is collected, the contribution
$f_\textmd{inc}$ leads to a limitation in the efficiency of the
setup. To quantify the error probability, we define
\begin{equation}
{\cal E} =
\frac{\int d \Omega \ f_\textmd{inc}(\Omega) }{\int d \Omega \
  f(\Omega)}.
\label{error.definition}
\end{equation}
As long as the number of excited atoms is small $n_x \ll N$,
this analysis can be easily generalized to the emission of states with  
many photons (\ref{multi.W.state}). One obtaines that the overall
error is bounded by $1 - \left(1 - {\cal E} \right)^{n_a + n_b}$.

The emission pattern can be obtained by studying the Heisenberg
equations of motion of the field operators. The calculation inolves
the study of the decay of the atomic state under collective effects
(see Appendix \ref{photon.distribution}). To simplify our analysis we ignore
the dipole pattern, in which case we get:
\begin{equation}
f(\Omega) =
\frac{1}{N}
\sum_{i,j = 1}^N
\langle
e^{-i (k_\textmd{L} {\bf n}_\Omega - {\bf k}_\textmd{L})({\bf r}_i - {\bf r}_j)}
\rangle
e^{ i {\bf k}_0 ({\bf r}^0_i - {\bf r}^0_j)  }
.
\label{spe.general}
\end{equation}
$\rr_j$ are the coordinate operators of the atoms, and thus
Eq. (\ref{spe.general}) allows us to describe fluctuations in
the position of the particles during the emission of light. 
In the following, we will show three different experimental
set--ups where our scheme can be implemented. In order to analyze
the performance in each of them, we first particularize the
above formula to three different situations which are directly
connected with those set--ups. We will focus on the angular width of
the forward--scattering cone, $\Delta \Omega$, which measures the
collimation of the emitted photons, and the error
probability, ${\cal E}$, as figures of merit.
Then we will introduce the possible
implementations and will use those formulas to specify the
conditions for them to correctly operate.

{\it (i) Fixed atomic positions.} 
In the case of a square lattice
of particles trapped in 3D (see Fig. \ref{general} (b)), the emission
pattern is given by
\begin{equation}
f_0(\Omega) =
\frac{1}{N} \hspace{-0.2cm}
\prod_{\alpha = x,y,z} \hspace{-0.2cm}
\frac{\sin^2( ( k_\LL \nn_\Omega^\alpha - (\kk_\LL \! + \! \kk_0)^\alpha)
  d_0 N_\alpha / 2)}
     {\sin^2( ( k_\LL \nn_\Omega^\alpha - (\kk_\LL \! + \! \kk_0)^\alpha) d_0 /2)} ,
\label{f0.fixed.positions}
\end{equation}
with $N_\alpha$ the number of atoms in each direction.
$f_0(\Omega)$ has a series of diffraction peaks, which are reduced to
a single one if $d_0 < \lambda/2$. In this regime, the emission is
centered in a cone with $\nn_\Omega$ in the direction of $\kk_\LL + \kk_0$.
Note that for simultaneous energy
and momentum conservation condition $|\kk_\LL + \kk_0| = k_\LL$ has to
be fulfilled. 
Since the positions of the atoms do not fluctuate, $f_0$ has only a
coherent contribution (${\cal E} = 0$), and the only limitation for
the effiency of the setup is the width of the emission cone, which scales in 3D like
$\Delta \theta_{\textmd{3D}} \approx 1/(N^{1/3} k_\LL d_0)$.
In the case of a chain of atoms (1D) momentum is conserved only along
the direction of the chain. Photon emission can be still
directed efficiently along the axis of the chain, in a cone whose
width scales like $\Delta \theta_{\textmd{1D}} \approx 1/\sqrt{N k_\LL d_0}$.

{\it (ii) Fluctuating atomic positions.}
Let us consider a lattice of atoms at temperature $T$, trapped by
independent harmonic potentials. The emission pattern is now the sum the of
two contributions,
\begin{equation}
f_\textmd{coh}(\Omega) = f_0(\Omega) g_T(\Omega), \hspace{0.2cm}
f_\textmd{inc}(\Omega) =  1 - g_T(\Omega) .
\label{emission.pattern.2}
\end{equation}
$g_T (\Omega) = e^{- ((k_\LL \nn_\Omega - \kk_\LL) {\bf \xi}_T)^2}$, and
$\xi_T$ is the vector whose components are the size of the position
fluctuations in each spatial direction, $(\xi_T^\alpha)^2 = x_0^2 ( 1
+ 2  n^\alpha_T)$, with $x^\alpha_0$ the size of the ground state in harmonic
potential, and $n^\alpha_T$ the number of motion quanta at $T$.
Light scattered into $f_\textmd{inc}$
represents an important fraction whenever $\xi_T^\alpha \gg d_0$.
In this case, the emission of light is centered around $\kk_\LL$,
since the uncertainty in the position of the particles averages out the
intial linear momentum $\kk_0$. The scaling of ${\cal E}$ in this
regime strongly depends on the dimensionality of the system. In
particular, in the case of a chain of atoms, ${\cal E}_\textmd{1D}
= d_0/\lambda$, whereas in the square 3D lattice, we get ${\cal
E}_\textmd{3D} \approx 12.6 (d_0/\lambda)^2 N^{-1/3}$.

{\it (iii) Statistical distribution of particles.}
Consider an ensemble of atoms (see Fig. \ref{general} (c)), which
move inside a square box of size $L$, such that their motion is
faster than their radiative decay, that is, their average velocity
$v$ is such that $v L \gg \Gamma$, with $\Gamma$ the emission rate.
This situation can be described
by assuming that the atoms are in a statistical distribution with
equal probability to be at any point in the box. The situation is
thus similar to that of a thermal state,
\begin{equation}
f_\textmd{coh} \left( \Omega \right) = N g_\textmd{box} (\Omega),
\hspace{0.2cm} f_\textmd{inc} \left( \Omega \right) = 1 -
g_\textmd{box}(\Omega), \label{emission.pattern.3}
\end{equation}
where
\begin{equation}
g_{\textmd{box}}(\Omega)  =
\prod_{\alpha = x,y,z}
\textmd{sinc}^2 \left( L ( k_L {\bf n}_\Omega^\alpha - {\bf
    k}_\LL^\alpha )
\right).
\label{statistical.distribution.box}
\end{equation}
Defining the average distance between particles like $d_0 = L
N^{-1/3}$, we find the same scalings of $\Delta \theta$ as in case
${\it (i)}$, and of $\cal E$, as in case ${\it (ii)}$.
Trapping schemes for atomic ensembles are simpler to realize but face
the difficulty that conditions for the directionality of photon
emission are more stringent. In the
case of a lattice of particles at fixed positions, forward--scattering is ensured
whenever condition $d_0 < \lambda/2$ is fulfilled. On the contrary, in
the case of atomic ensembles, the incoherent contribution
$f_\textmd{inc}$ has to be small enough such that ${\cal E} \ll 1$,
which implies $d_0 \ll \lambda$ in 1D, or, alternatively, a number of
particles large enough in 3D.

Now we introduce three experimental set--ups where our scheme can
be implemented. In the Appendix we show how to create the
atomic states that we are considering here.

\begin{figure}[h]
\center
\resizebox{\linewidth}{!}{\includegraphics{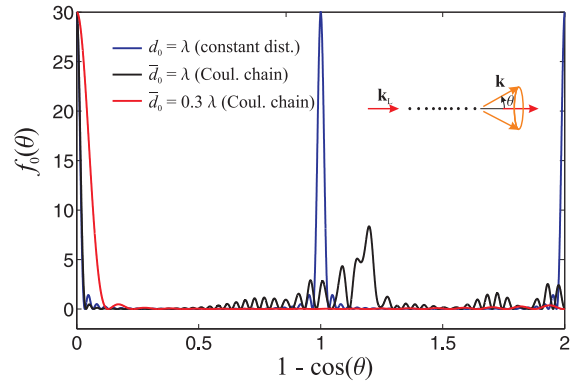}}
\caption{Probability of photon emission from an ion chain with
  $N =$ 30 ions initially in a $W$--state. The blue line corresponds
  to a chain with equally spaced ions with two diffraction
  peaks. Black and red lines corresponds to an ion Coulomb chain, in
  which ions are in an overall trapping potential and thus are not
  equally spaced.
 However, in the case that the average distance, $\bar{d}_0$ is small
  enough, light is also preferentially emitted in the
  forward--scattering direction.}
\label{ions}
\end{figure}

{\it  Trapped ions.} This system is ideally suited to create
collective states like (\ref{multi.W.state}), 
as was demonstrated recently in ref. \cite{Häffner}.
Most usually ions are arranged in chains, such that we deal with
the 1D situation discussed above. Even though trapped ions are not
equally spaced, under the condition $\bar{d}_0 < \lambda /2$,
with $\bar{d}_0$ the average distance, we still get light emission
in the forward--scattering cone only, see Fig. \ref{ions}.
Considering two different internal levels, which can correspond to
different states in an hyperfine multiplet, states such as those
defined by Eq. (\ref{multi.W.state}) can be created by a number of
quantum operations that scales linearly with the number of ions
$N$ (see \ref{MPS}). For example, the state $1/\sqrt{2} \left( |
{\bf 0} ,2 \kk_L \rangle + | 2 \kk_L, {\bf 0} \rangle \right) $,
would emit two photons in the forward and backward directions
along the chain axis, entangled in polarization. The main
difficulty for the implementation of this idea with ions lies on
the fact that ion--ion distances are usually in the range of a few
$\mu m$, and thus condition $d_0 < \lambda/2$ is not fulfilled
when considering optical wavelengths. A way out of this problem is
to use optical transitions which lie in the range of $\lambda
\gtrsim 5 \mu m$, which can be found in ions such as Hg$^+$,
Ba$^+$, or Yb$^+$ \cite{note.ions}.

{\it Cold atoms in optical lattices.}
By using optical lattices we fulfill the need of placing atoms at
interparticle distances comparable to optical wavelengths,
since potential wells in a
standing--wave are indeed separated by
$d_0 = \lambda_{\textmd{sw}}/2$, with $\lambda_{\textmd{sw}}$, the
wavelength of the counterpropagating lasers.
By using an optical transition such that
$\lambda > \lambda_{\textmd{sw}}$, we are in the regime in which
light emission is focused into a single Bragg peak.
Although one could think of peforming quantum gates between
ultracold neutral atoms to generate collective atomic states \cite{Jaksch},
this procedure faces the difficulties of quantum computation in this system, like
for example how to achieve single atom addressability.
More efficiently, one could avoid the use of quantum gates by using
the dipole--blockade mechanism with Rydberg atoms, which allows us to
generate W-states, as well as states which emit Fock states with a
number $M$ of photons  \cite{LukinRydbergatoms} (see Appendix \ref{Rydberg}).

{\it Atomic ensembles at room temperature.} The very same
techniques which can be applied to Rydberg atoms in an optical
lattice can also be used in the case of hot ensembles.
On the one hand, this setup has the advantage that atoms do not need to be
cooled and placed in an optical lattice. On the other hand, it can be
described by a statistical distribution of particles, and thus
suffers from the fact that high efficiency in the release of
photons is achieved under more severe conditions of particle
density and atom number, as discussed above. However, densities which
are high enough to fulfill the requirement ${\cal E} \ll 1$ have been
recently reported in \cite{Heidemann}.

In conclusion, we have proposed to use current techniques for
quantum engineering to generate atomic multipartite entangled
states which can be efficiently mapped into photonic states. Our
proposal relies on the release of spin--wave like excitations into
a given spatial direction by means of interference effect, and can
be implemented with trapped ions, atoms in optical lattices, and
atomic ensembles at room temperature.

This work was supported by E.U. projects (SCALA and CONQUEST), and the
Deutsche Forschungsgemeinschaft.

\appendix

\section{Creation of collective atomic states in a chain of atoms}
\label{MPS}
Entangled states of the form (\ref{multi.W.state}) and their
linear combinations can be generated in a chain of particles, for
example, of trapped ions, by means of a limited number of quantum
operations. To demonstrate this, we first show that they can be
written as Matrix Product States with a small bond dimension $D$,
i.e. they can be written as
\begin{equation}
| \Psi \rangle = \sum_{i_1,\dots,i_N} \langle \Phi_\textmd{F} |
V_{[N]}^{i_N} \dots V_{[1]}^{i_1} | \Phi_\textmd{I} \rangle | i_1
\rangle \dots | i_N \rangle , \label{MPS}
\end{equation}
In (\ref{MPS}), the indices $i_j = g, s_a, s_b$, and
$V^{i_j}_{[j]}$ are $D \times D$ matrices acting on an auxiliary
$D$--dimensional Hilbert space. $D$ is given by the number of
states which appear in the singular value decomposition (s.v.d.) of 
$| \Psi \rangle$ at any site in the chain \cite{Vidal}. As it is shown in
\cite{MPS.sequential.generation}, the state (\ref{MPS}) can be
prepared by performing $N$ gates which act on $[\log_2D]+1$
qubits. Thus, as long as $D$ is independent of $N$, the number of
gates to be applied scales linearly with the total number of
atoms.

To evaluate $D$, consider first the case of a state like
(\ref{multi.W.state}) with atoms excited in level $s_a$ only, 
and a partition of the chain in two parts $L$ and $R$. 
We get $n_a + 1$ states in the s.v.d. with respect to this partition,
 which correspond to states with
a number of excited atoms in part $L$, ranging from $0$ to $n_a$.
This result is easily generalized to a linear combination
of $M$ states of the form (\ref{multi.W.state}), in which case we get
$D = M(n_a + 1)(n_b+1)$. For example, an entangled state of the form:
\begin{equation}
| \Psi \rangle =
\frac{1}{\sqrt{2}}
\left( | {\bf k}^{n_a = 1}, {\bf q}^{n_b = 1} \rangle
+      | {\bf q}^{n_a = 1}, {\bf k}^{n_b = 1} \rangle \right) ,
\label{2.spin.waves}
\end{equation}
has $D=8$.

\section{Quantum state engineering with Rydberg blockade}
\label{Rydberg}
Interactions between excited atomic states, like those that take
place in Rydberg atoms, can be used to the create
the states defined by Eq. (\ref{2.spin.waves}).
This can be achieved in a single experimental step, without the need
for quantum gates, if the proper configuration of atomic interactions
is chosen.
As an example, consider the 3 level configuration shown in
Fig. \ref{general} (a), and interactions between excited states such that
atoms in levels
$| s_a \rangle$, $| s_b \rangle$, interact strongly only if they are in the
same excited state, that is, $U_{aa} = U_{bb} = U$, but  $U_{ab}=0$.
We apply two lasers with wavectors $\kk_{1,2}$ and Rabi
frequencies $\Omega_{1,2}$,
detuned with respect to the $| g \rangle$ -- $| s_{a,b} \rangle$ transition,
such that $\Delta_1 = - \Delta_2 = \Delta$. If condition $\Delta_{1,2} \gg
\Omega_{1,2}$ is fulfilled, then the lasers induce a two--photon
transition with Rabi frequency $\Omega_\textmd{eff} = \Omega_1
\Omega_2 / \Delta$. Furthermore, if $\Omega_\textmd{eff} \ll U$,
states with two atoms in the same excited state are not populated.
Under these conditions there are two possible excitation channels,
depicted in Fig. \ref{Rydberg}, which give rise to the linear
combination (\ref{2.spin.waves}).
\begin{figure}[h]
\center
\resizebox{\linewidth}{!}{\includegraphics{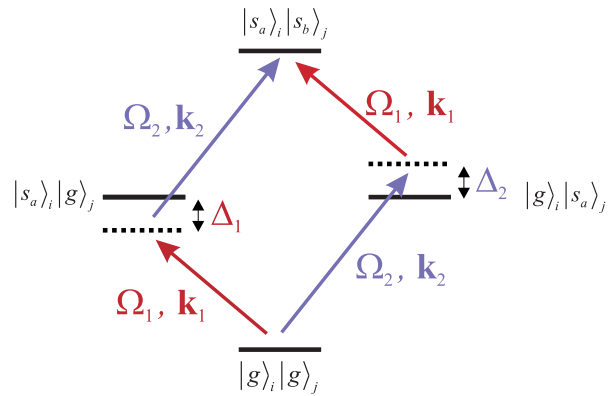}}
\caption{Lasers and level configuration for the creation of atomic
  entangled states which emit pairs of photons entangled in polarization.}
\label{Rydberg}
\end{figure}
\section{Calculation of the photon distribution}
\label{photon.distribution}
We consider for simplicity the lambda
configuration depicted in Fig. \ref{general} (a), considering a single
excited state $| s \rangle$, and a single auxiliary level $| e \rangle$. The
interaction of the quantized electromagnetic field with the ensemble
of atoms, after the adiabatic elimination of level $| e \rangle$,
is described by
\begin{eqnarray}
H_{\textmd{I}} &=&
\sum_{j,{\bf k},\lambda} g_{{\bf k} \lambda}
\left( \sigma^\dagger_j  a_{{\bf k}, \lambda}
e^{i ( {\bf k} - {\bf k}_L) {\bf r}_j + i \omega_L t}  + h.c. \right),
\nonumber \\
g_{{\bf k}, \lambda} &=& \frac{\Omega_L}{2 \Delta}
\sqrt{\frac{\hbar\omega_k}{2 \epsilon_0 V}}
\left( \epsilon_{{\bf k} \lambda} \cdot {\bf d}_{ge} \right) ,
\label{interaction.hamiltonian}
\end{eqnarray}
$\sigma_j$ refers to the $| g \rangle$ -- $| s \rangle $ atomic
transition, $\Omega_L$ and $\kk_\LL$ are the Rabi frequency and
wave--vector of the classical field, respectively,
$\omega_k$ is the photon energy, $\epsilon_{\kk,\lambda}$
are the polarization vectors, and ${\bf d}_{ge}$ is the dipole matrix
element for the $| g \rangle$ -- $| e \rangle$ transition.

The probability of photon emission is proportional to the
the diagonal elements of the one--photon density matrix, which are
obtained from the Heisenberg equation of motion for the field
operators,
\begin{eqnarray}
&& \langle a^\dagger_{\bf k} a_{{\bf k}}  \rangle
= \frac{1}{N}
\sum_\lambda  g_{\kk, \lambda}^2
%\nonumber \\
%&& \hspace{0.5cm}
\int_0^\infty \! \! \! \!  d \tau_1  d \tau_2
e^{ - i (\omega_k-\omega_L) (\tau_1 - \tau_2)}
\nonumber \\
&& \hspace{1.5cm}
\sum_{ij}
\langle
e^{-i ({\bf k} - {\bf k}_L) ({\bf r}_i - {\bf r}_j)}
\rangle \langle \sigma^\dagger_i (\tau_1) \sigma_j (\tau_2) \rangle.
\label{one.photon}
\end{eqnarray}
Since we are interested in the conditions for momentum conservation
due to interference effects, we consider the following atomic initial
state,
\begin{equation}
| \kk_0 \rangle = \sigma^\dagger_{\kk_0} | 0 \rangle,
\hspace{0.2cm} \sigma^\dagger_{\kk_0} = \frac{1}{\sqrt{N}} \sum_{j=1}^N e^{-i
\kk_0 \rr^0_j} \sigma^\dagger_j .
\label{W.state}
\end{equation}
The emission pattern depends thus on the two--time atomic correlation
function, which in turns can be evaluated by means of a master
equation which describes the decay of the atomic levels. In the case
of the initial atomic state $\kk_0$ (\ref{W.state}), fixed atom
positions, and neglecting
boundary effects, this correlation function can be evaluated exactly,
\begin{equation}
\langle \sigma_i^\dagger(\tau_1) \sigma_j(\tau_2)\rangle
= e^{- \Gamma_\kk (\tau_1 + \tau_2)/2}
e^{i \kk_0 (\rr^0_i - \rr^0_j)} ,
\label{2.time.correlator}
\end{equation}
where we have neglected an energy shift due to dipole--dipole
interactions. Integrating
(\ref{one.photon}) over the absolute value of $\kk$ yields the
probability of photon emission,
\begin{equation} 
I(\Omega) = \bar{I}(\Omega) f(\Omega).
\end{equation}
$\bar{I}(\Omega)$ is the dipole pattern,
\begin{equation}
\bar{I}(\Omega) = 
\frac{3}{8 \pi}
\frac{\Gamma}{\Gamma_{\kk_0}}
\left( 1 - \left( {\bf n}_{eg} {\bf n}_\Omega \right)^2
\right), 
\label{dipole.pattern}
\end{equation}
where $\Gamma$ is the single atom radiative decay rate,  $\Gamma_{\kk_0}$ is the
collective decay rate,
${\bf n}_{eg}$ is the unit vector of the atomic transition,
and $\nn_\Omega$ is a unit vector in the direction defined by the
solid angle $\Omega$.
The factor $f(\Omega)$ in $I(\Omega)$ describes the 
interference between the emission from different atoms, and is given
by Eq. (\ref{spe.general}).

Below we deduce the master equation which leads to
(\ref{2.time.correlator}) and we sketch its solution in the case of
collective states with a single excited atom.

\section{Master Equation}
\label{master}
The master equation for the reduced density matrix of the internal
levels, which describes the ratiative decay of a
set of atoms under the coupling to the quantized radiation field given
by Eq. (\ref{interaction.hamiltonian}), is
\begin{eqnarray}
\partial_t \rho &=&
\sum_{i,j} \frac{\Gamma_{ij}}{2}
\left( 2 \ \sigma_i \rho \sigma^\dagger_j
- \sigma^\dagger_i \sigma_j \rho - \rho \sigma^\dagger_i \sigma_j
\right) \nonumber \\
&+&
\frac{i}{2} \sum_{ij} G_{ij} [\sigma^\dagger_i \sigma_j, \rho],
\label{master.dipole}
\end{eqnarray}
where the coupling constants depend on
\begin{eqnarray}
J_{ij} &=& \int_0^\infty d \tau g_{ij}(\tau) e^{-i \omega_L \tau} =
\nonumber \\
&&  \hspace{-1.2cm} = \! g^2 \! \int_0^\infty
\sum_{{\bf k}, \lambda}
\frac{\hbar \omega_k}{2 \epsilon_0 V}
\left( \epsilon^\lambda_{\bf k}  {\bf d}_{ag} \right)^2
e^{i (\omega_k - \omega_L) \tau + i ({\bf k} - {\bf k}_L){\bf r}} ,
\label{master.dipole.J}
\end{eqnarray}
in the following way:
\begin{eqnarray}
\Re(J_{ij}) &=& \frac{1}{2} \Gamma_{ij}, \nonumber \\
\Im(J_{ij}) &=& \frac{1}{2} G_{ij}.
\label{master.dipole.Gamma}
\end{eqnarray}
The master equation (\ref{master.dipole}) can be solved for the
particular case of an initial state (\ref{W.state}) by noticing that
the evolution of $\rho$ is closed within the subspace spanned by the
states $| \kk_0 \rangle$, $| 0 \rangle$. This fact can be easily
proved by direct substitution of
$\rho(0) = | \kk_0 \rangle \langle \kk_0 |$ in
Eq. (\ref{master.dipole}), which yields the following evolution for
the atomic density matrix:
\begin{equation}
\rho(t) = e^{-\Gamma_{\kk_0} t} | \kk_0 \rangle \langle \kk_0 | +
\left( 1 - e^{-\Gamma_{\kk_0} t} \right) | 0 \rangle \langle 0 |,
\label{rho.evolution}
\end{equation}
where the collective decay rate $\Gamma_{\kk_0}$ is just the Fourier
transform of the coupling constants in the master equation,
\begin{equation}
\Gamma_{\kk_0} = \sum_j \Gamma_{i,j} e^{i \kk_0 (\rr^0_i - \rr^0_j)} .
\label{foruier.transform}
\end{equation}
A similar result holds for nondiagonal elements of $\rho(t)$, which
together with the quantum regression theorem yields the evolution of
the atomic correlation function (\ref{2.time.correlator}).

\end{document}